\documentclass[letterpaper, 10 pt, conference]{ieeeconf}  
\IEEEoverridecommandlockouts   
\usepackage{graphicx}
\usepackage{amsmath}
\usepackage{amssymb} 
\usepackage{xcolor}
\usepackage{soul}
\setstcolor{red}

\usepackage[hidelinks]{hyperref}
\usepackage[utf8]{inputenc}
\usepackage[small]{caption}
\usepackage{booktabs}
\usepackage[switch]{lineno}
\usepackage{algorithm}
\usepackage{algorithmic}
\usepackage{tikz}
\usepackage{multirow}
\usepackage{subcaption}

\usepackage{amsthm}
\newtheorem{thm}{Theorem}
\newtheorem{definition}[thm]{Definition}
\usepackage{balance}
\usepackage{bbding}
\usepackage{threeparttable}
\usepackage{multirow}
\usepackage{array}
\usepackage{bm}

\definecolor{darkgreen}{rgb}{0.0, 0.5, 0.0}
\definecolor{lightred}{rgb}{1.0, 0.8, 0.8}
\definecolor{lightgreen}{rgb}{0.8, 1.0, 0.8}
\definecolor{lightblue}{rgb}{0.8, 0.8, 1.0}

\newcommand{\redX}{\textcolor{red}{\scalebox{0.7}{\textbf{\XSolidBrush}}}}
\newcommand{\greenCheck}{\textcolor{darkgreen}{\scalebox{0.7}{\textbf{\Checkmark}}}}

\usepackage{pifont}

\title{\LARGE \bf 
Mitigating Side Effects in Multi-Agent Systems Using Blame Assignment
}

\author{Pulkit Rustagi$^{1}$ and Sandhya Saisubramanian$^{1}$
\thanks{$^{1}$The authors are with the Collaborative Robotics and Intelligent Systems (CoRIS) Institute, Oregon State University, Corvallis OR 97331, USA
        {\tt\small \{rustagip, sandhya.sai\}@oregonstate.edu}}}

\begin{document}

\maketitle
\thispagestyle{empty}
\pagestyle{empty}

\begin{abstract}
When independently trained or designed robots are deployed in a shared environment, their combined actions can lead to unintended negative side effects (NSEs). To ensure safe and efficient operation, robots must optimize task performance while minimizing the penalties associated with NSEs, balancing individual objectives with collective impact. We model the problem of mitigating NSEs in a cooperative multi-agent system as a bi-objective lexicographic decentralized Markov decision process. We assume independence of transitions and rewards with respect to the robots' tasks, but the joint NSE penalty creates a form of dependence in this setting. To improve scalability, the joint NSE penalty is decomposed into individual penalties for each robot using credit assignment, which facilitates decentralized policy computation. We empirically demonstrate, using mobile robots and in simulation, the effectiveness and scalability of our approach in mitigating NSEs.
\end{abstract}

\section{INTRODUCTION}
In many real-world environments, the actions of individual robots, even when operating independently, can have significant and often unintended consequences on the broader environment and the collective system behavior. Traditional approaches to agent design and training focus on optimizing performance in isolation, prioritizing task completion while overlooking potentially harmful effects their actions can collectively produce, such as \emph{negative side effects} (NSEs). In multi-agent systems, NSEs are the unintended and undesirable outcomes of collective system behavior, caused by the incompleteness of models used for decision making~\cite{alizadeh2022considerate,saisubramanian2021multi,saisubramanian2022avoiding}. While the NSEs are difficult to identify before deployment, mitigating them is crucial for overall system efficiency and safety.

This paper focuses on mitigating NSEs in cooperative multi-agent settings where the robots produce no (or negligible) NSEs when executing their policy in isolation, but their joint policy results in NSEs. Consider warehouse robots that optimize moving shelves between two locations. Each robot's model provides the necessary information, including reward and transition dynamics, to complete its task optimally. The models lack information about the effects of robots' joint actions in the environment, such as a narrow corridor being blocked for human access when multiple robots simultaneously move large shelves through it. Thus, even when the robots are adept at their tasks and produce no NSEs when acting in isolation, their joint actions are undesirable. 
\begin{figure}[t]
    \centering
   \includegraphics[width=0.98\linewidth]{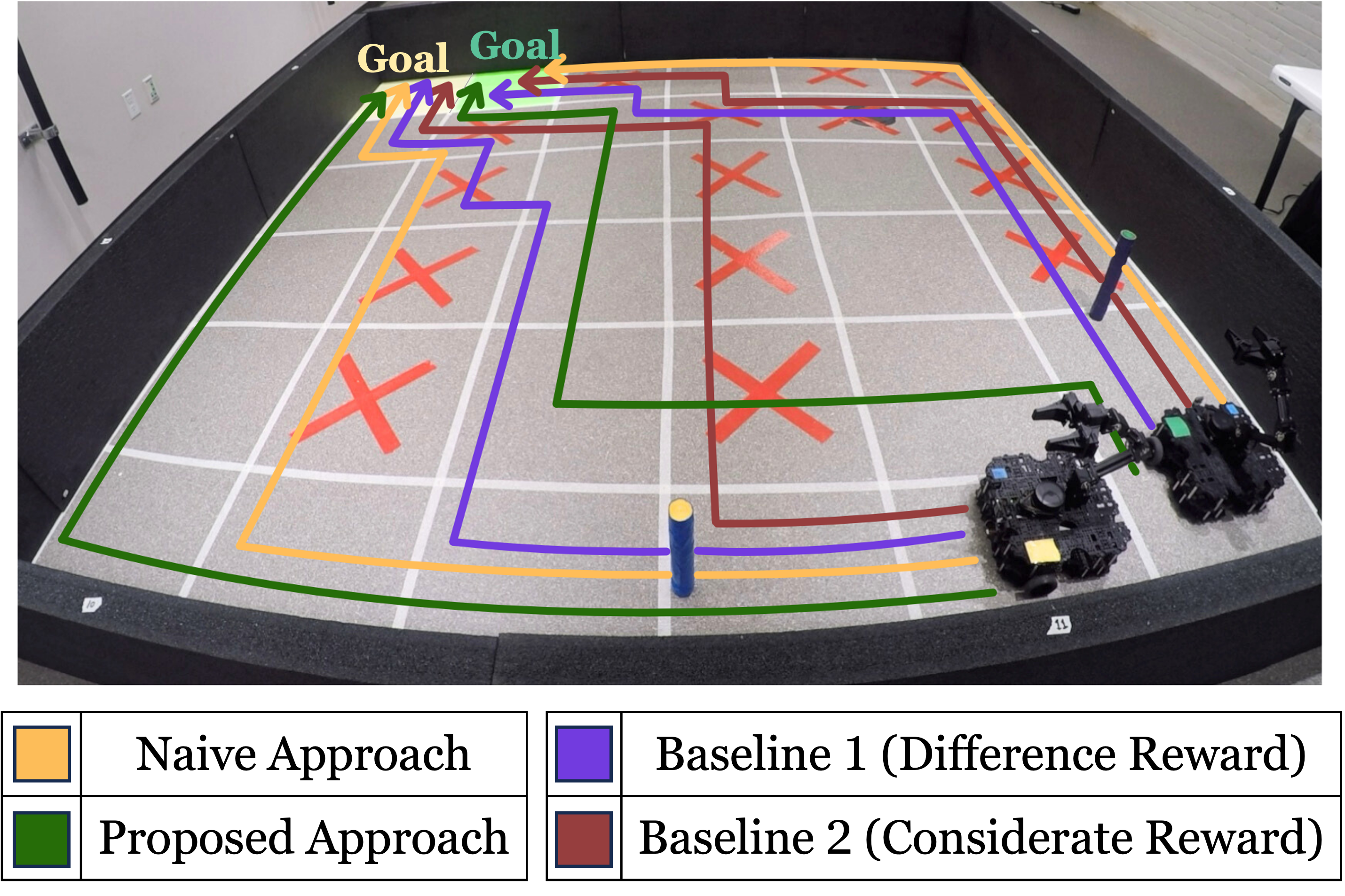}
    \caption{
    Comparison of the paths taken by two TurtleBots performing delivery tasks in our indoor setup. The robots, initially unaware about the side effects of their actions, receive a joint penalty when one or both are in NSE states marked by X. The robots must update their behavior to complete tasks while mitigating NSEs. 
    }
    \label{fig:experimental_setup}
\end{figure}

Mitigating NSEs in multi-agent settings is challenging because NSEs and corresponding penalties are defined over joint actions, creating inter-agent dependencies that did not exist for task completion. Further, the computational complexity of mitigating NSEs, without significantly affecting task completion, increases with the number of agents in the system. Prior research on NSEs primarily target single-agent settings~\cite{klassen2022planning,krakovna2020avoiding,saisubramanian2021multi,saisubramanian2022avoiding,Zhang_Durfee_Singh_2018} or treat other agents as part of the environment~\cite{alizadeh2022considerate}. These techniques do not apply to multi-agent settings as they ignore the agent interactions that produce NSEs. A recent approach uses distributed constraint optimization with Q-learning to mitigate multi-agent NSEs~\cite{choudhury2024minimizing}. However, it is not scalable since distributed constraint optimization is intractable for large problems~\cite{fioretto2018distributed}.

\begin{figure*}[t]
    \centering
    \includegraphics[scale=0.2,trim={7.3cm 0 10.5cm 0},clip]{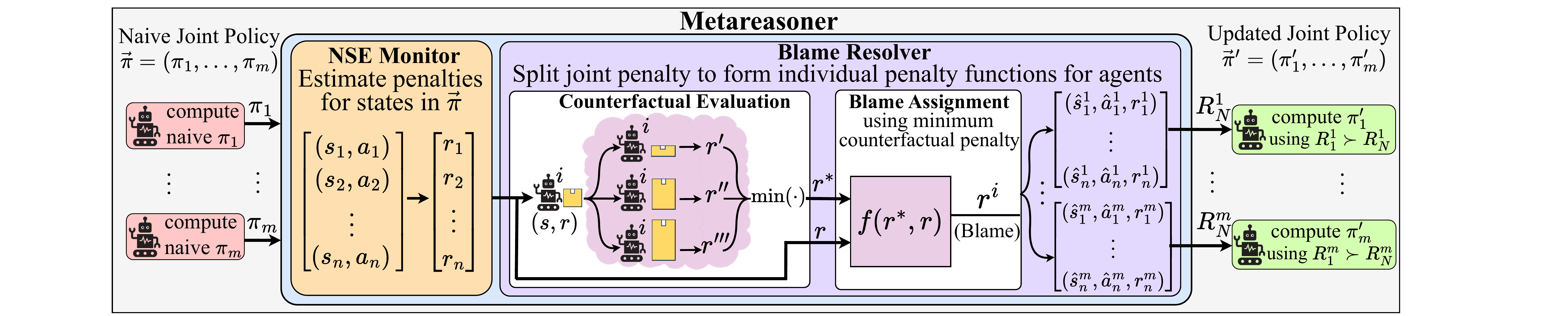}
    \caption{Overview of our solution approach. Agents independently compute policies to complete tasks described by $R_1^i$ (Naive policy). The NSE Monitor computes the NSE penalty for the joint policy $\vec{\pi}$. The Blame Resolver assigns a blame value for each agent, by evaluating counterfactual scenarios specific to each agent, as illustrated with warehouse robots handling different-sized boxes. Individual penalty functions $R^i_N$ are derived for each agent, based on the estimate blame.
    Agents then recompute their policies by solving the bi-objective problem with $R_1^i \succ R_N^i$, where $\succ$ denotes preference ordering over the objectives and their associated reward functions.
    }
\label{fig:detailed_approach}
\end{figure*}

We formulate the problem of mitigating NSEs in a multi-agent system as a decentralized Markov decision process (DEC-MDP) with two objectives and a \emph{lexicographic ordering} over them. A 
lexicographic formulation is an intuitive way to model problems with an inherent ordering over objectives~\cite{rentmeesters1996theory,wray2015multi}. The \emph{primary objective} for each agent is to optimize its assigned task. We assume independence of transition and rewards with respect to the agents' assigned tasks~\cite{becker2004solving}. That is, each agent's reward for task completion is determined by its local state and actions and the overall reward for the system is the sum of individual agent's rewards, and agents' state transitions are affected only by their own actions. The \emph{secondary} objective is to minimize the NSEs. Agents have no prior knowledge of NSEs and only observe the \emph{joint penalty}---a global value based on \emph{all} agents' actions. While they can independently compute policies for their tasks, mitigating NSEs requires addressing the dependence induced by the joint penalty (Fig.~\ref{fig:experimental_setup}). 

We present a \emph{metareasoning}~\cite{carlin2012bounded,SZmetareasoning11} approach to detect and mitigate NSEs. The metareasoner is a centralized entity that monitors the agents' behaviors and has two components: (1) \textit{NSE Monitor} that estimates the NSEs associated with agents' \textit{joint policy}, and outputs the corresponding penalty, and (2) a \textit{Blame Resolver} that decomposes the joint penalty into individual agent penalties, using our algorithm \textit{Reward Estimation using Counterfactual Neighbors (RECON)}. It is assumed that the NSE monitor has access to a model of NSEs and the associated penalty, either provided as part of its design or acquired through human feedback. The blame resolver, using RECON, performs blame (credit) assignment to determine each agent's relative contribution towards the NSE, based on which the joint NSE penalty is decomposed into local penalties for each agent~\cite{bilbao2000shapley,nguyen2018credit,proper2012modeling}. This decomposition facilitates augmenting the model of each agent with NSE information via a penalty function, thereby enabling decentralized policy computation to mitigate NSEs. 

Our solution framework uses a four-step approach to mitigate NSEs (Fig.~\ref{fig:detailed_approach}): (1) the agents first calculate optimal policies to complete their assigned tasks (referred to as naive policies); (2) the NSE monitor estimates the NSE penalty associated with the joint policy of the agents; (3) the blame resolver then decomposes the joint penalty into individual penalties for each agent, using blame (credit) assignment; and (4) the agents recompute their policies by solving a decentralized, bi-objective problem, with the prescribed reward for their task and the estimated local penalty for NSE, using lexicographic value iteration~\cite{wray2015multi}. Our experiments using mobile robots and in simulation demonstrate the efficiency and scalability of our approach in mitigating NSEs by updating the policies of a subset of agents in the system.

\begin{table*}[t]
\centering
\caption{Overview of different credit assignment techniques for mitigating NSEs. A technique is scalable if it is computationally inexpensive to generate counterfactuals for problems with a few hundreds of agents. ``$-$'' indicates that the approach cannot be applied directly but can be modified to meet the requirements of our setting. }
\small
\begin{threeparttable}
\begin{tabular}{|c|c|c|c|c|}
\hline
\multirow{2}{*}{\centering\arraybackslash Credit Assignment Technique} &
\multirow{2}{*}{\centering\arraybackslash { Scalable}} &
\multirow{2}{*}{\centering\arraybackslash \raisebox{6pt}{Supports decentralized}} &
\multirow{2}{*}{\centering\arraybackslash \raisebox{6pt}{Compatible with}} &
\multirow{2}{*}{\centering\arraybackslash \raisebox{6pt}{Can generate NSE-}} \\ 
& & \raisebox{3pt}{planning} & \raisebox{3pt}{heterogeneous agents}& \raisebox{3pt}{specific counterfactuals}\\ \hline
$\text{Shapley Value}$ & \redX & \greenCheck & \greenCheck & \redX \\ \hline
$\text{Difference Reward}$ & \greenCheck & \greenCheck & \greenCheck & $-$ \\ \hline
$\text{Action Not Taken}$& \greenCheck & \greenCheck & \greenCheck & \redX \\ \hline
$\text{D{++}}$ & \redX & \greenCheck & \greenCheck & \redX\\\hline
$\text{Wonderful Life Utility}$& \redX & \greenCheck & \redX & \redX \\ \hline
$\text{Value Decomposition}$& \redX & \redX & \greenCheck & \redX \\ \hline
$\text{Our Approach}$& \greenCheck & \greenCheck & \greenCheck & \greenCheck \\ \hline
\end{tabular}
\end{threeparttable}
\label{tab:credit_assgn}
\end{table*}

\section{BACKGROUND} 
\noindent \textbf{Decentralized Markov Decision Processes (Dec-MDPs)~} are widely used to model decentralized multi-agent decision-making problems~\cite{goldman2004decentralized}. A Dec-MDP is defined by the tuple $\langle \mathcal{A},S,A,T,R \rangle$ with $\mathcal{A}$ denoting the finite set of $k$ agents in the system; $S\!=\!\hat{S}_1\!\times\!\ldots\!\times\!\hat{S}_k$ denoting the joint state space, where $\hat{S}_i$ denotes the state space of agent $i$; $A=\hat{A}_1\!\times\ldots\!\times\!\hat{A}_k$ denoting the joint action space, where $\hat{A}_i$ denotes agent $i$'s actions; $T:S\times A\times S\rightarrow [0,1]$ denoting the transition function; and $R$ denoting the reward function. We use $s_{-i}$ to denote joint state excluding agent $i$. A joint policy $\vec{\pi}\!=\!(\pi_1,...,\pi_k) $ is a set of policies, one for each agent in the system.  A Dec-MDP with \emph{transition independence and reward independence}~\cite{becker2004solving} is a class of problems in which agents operate independently but are tied together through a reward structure that depends on all of their execution histories:
\[ T(\hat{s}_i'|{s}, {a}, {s}'_{-i}) = T_i(\hat{s}_i'|\hat{s}_i,\hat{a}_i), \forall i \in \mathcal{A},\] 
\[ R({s}, {a}, {s}') = \sum_{i\in\mathcal{A}} R_i(\hat{s}_i,\hat{a}_i,\hat{s}_i').\]
A transition and reward-independent Dec-MDP can be solved as $k$ single agent MDPs~\cite{goldman2004decentralized}.

\noindent \textbf{Lexicographic MDP (LMDP)~} LMDPs are particularly convenient to model problems with potentially competing objectives and an inherent lexicographic ordering over them, such as ours where task completion is prioritized over NSE mitigation~\cite{saisubramanian2021multi,wray2015multi}. 
An LMDP is denoted by the tuple $M = \langle S,A,T,\pmb{\mathrm{R}},o\rangle$ with finite set of states $S$, finite set of actions $A$, transition function denoted by $T\!:\!S\!\times\!A\times\!S\!\rightarrow\![0,1]$ and a vector of reward functions $\pmb{\mathrm{R} }\!=\![R_1,...,R_k]^T$ with $R_i\!:\!S\!\times\!A\!\rightarrow\!\mathbb{R}$, and $o$ denotes the strict preference ordering over the $k$ objectives. 
The set of value functions is denoted by $\pmb{V}\!=\![V_1,...,V_k]^T$, with $V_i$ corresponding to $o_i$,\\[-4pt]
\[ \pmb{V}^{\pi}(s)\!=\!\pmb{\mathrm{R}}(s,\pi(s)) + \gamma \sum_{s'\in S}T(s,\pi(s),s')\pmb{V}^{\pi}(s'),\forall s \in S.\]\\[-4pt]
A slack $\pmb{\Delta}\!=\!\langle \delta_1,..., \delta_k\rangle$ with $\delta_i\!\geq\!0$, denotes the acceptable deviation from the optimal expected reward for objective $o_i$ so as to improve  the lower priority objectives. Objectives are processed in the lexicographic order. The set of restricted actions for $o_{i+1}$ is $ A_{i+1}(s)\!=\!\{a\!\in\!A \vert \max \limits_{a'\in A_i} Q_i(s,a') - Q_i(s,a) \leq \eta_i\}$, where $\eta_i\!=\!(1-\gamma)\delta_i$,$\gamma \in [0,1)$. Refer~\cite{wray2015multi} for a detailed background on LMDP.

\noindent \textbf{Credit Assignment~} It is a popular approach to measure the contribution of an agent to team performance so as to convert a joint reward into individual agent rewards~\cite{nguyen2018credit}. Difference Reward~\cite{proper2012modeling} and its variants such as D++~\cite{rahmattalabi2016d++}, Wonderful Life Utility~\cite{nguyen2018credit}, perform credit assignment by comparing the joint rewards before and after the agent is removed from the system. Another type of credit assignment uses Shapley value to compute a value function for each agent by considering all combinations of possible agent interactions~\cite{bilbao2000shapley,meng2012core,shapley1953value,sundararajan2020many}. Table~\ref{tab:credit_assgn} summarizes the characteristics of different credit assignment techniques. While some of these techniques assess individual contributions, they do not support decentralized planning due to their reliance on other agents' policies. The existing methods calculate counterfactuals by considering all state features, including those that are not associated with NSEs, which leads to incorrect blame assignment for NSEs.

\section{PROBLEM FORMULATION}
\noindent \textbf{Problem Setting.~} Consider $m$ agents  independently performing their assigned tasks which is their primary objective $o_1\!=\!\{o_1^1,...,o_1^m\}$. Each agent operates based on an MDP $\hat{M}$ that contains all the information necessary to optimize $o_1$ but lacks information about the joint effects of agents' actions when it is unrelated to task completion. Due to the limited fidelity of $\hat{M}$, \emph{negative side effects} (NSEs) occur when agents execute their policies simultaneously. A meta-level process, \emph{metareasoner}, monitors and controls the agents' performance (object-level process)~\cite{svegliato2022metareasoning,SZmetareasoning11}, by providing a penalty if agents' actions produce side effects. The NSE penalty is determined by a function $R_N$ which is known to the metareasoner (either by design or learned using human feedback) but unknown to the agents. 

We assume the following about the NSEs: (1) agents acting in isolation produce no (or negligible) NSEs, but their joint actions produce NSEs that must be mitigated; (2) agents have no prior knowledge about NSEs of the joint actions except for the penalty assigned by the metareasoner; and (3) NSEs are undesirable but do not hinder task completion. We address the problem of mitigating NSEs by augmenting the agents' models with secondary reward functions that correspond to NSE penalties. We target settings where the completion of the agents' assigned tasks ($o_1$) are \emph{prioritized} over minimizing NSEs ($o_2$), $o_1\!\succ\!o_2$.

\noindent\textbf{MASE-MDP.~} The problem of mitigating NSEs in a cooperative multi-agent system is formulated as a Dec-MDP with two objectives and a lexicographic ordering over them.

\begin{definition} A multi-agent side effects MDP (MASE-MDP) is a bi-objective Dec-MDP with lexicographic ordering over the objectives, denoted by $M = \langle \mathcal{A}, S, A, T, \bm{R},o\rangle$, where:
    \begin{itemize}
        \item $\mathcal{A} = \{1,\hdots,m\}$ is a finite set of agents in the system; 
        \item $S = \hat{S}_1 \times \ldots \times \hat{S}_m$ is the joint state space;
        \item $A = \hat{A}_1 \times \ldots \times \hat{A}_m$ is the joint action space;
        \item $T:S\times A \times S \rightarrow [0,1]$ is the transition function; 
        \item $\bm{R}\!=\![R_1, R_N]$ is the reward function with $R_1\!:\!S\!\times\!A\!\rightarrow\!\mathbb{R}$ denoting the reward for task performance and $R_N\!:\!S\!\times\!A\!\rightarrow\!\mathbb{R}$ denoting the penalty function for NSEs; and
        \item $o=[o_1,o_2]$ denotes the objectives, where $o_1$ is the primary objective denoting agents' assigned tasks and $o_2$ is minimizing NSEs, with $o_1\succ o_2$. 
    \end{itemize}
\end{definition}

A MASE-MDP is characterized by independence of transition function and task completion reward $R_1$, meaning each agent's transitions and task reward depend only on its local state and action and is independent of other agents. This allows each agent to independently compute a policy for its assigned task~\cite{becker2004solving,goldman2004decentralized}. However, the agents incur a \emph{joint penalty} for the NSEs, which introduces a form of inter-agent dependence and prevents decentralized computation of individual policies. Section~\ref{sec:blame_assgn} describes an approach to decompose joint penalties into individual penalties, thereby facilitating decentralized planning. 

\noindent \textbf{Local and Global State Features.~ }  We consider a factored state representation. Let $F$ denote the set of features in the environment, which are categorized into \emph{local features} $F_l$ and \emph{global features} $F_g$, $F\!=\!F_l\cup\!F_g$. The local features of an agent $i$ are denoted by $F^i_l$. Local features are agent-specific features that are controlled by the agent's actions and affect its performance (e.g. $x,y$ location). Global features are shared among agents and denote the overall state of the system. They are further divided into static and dynamic global features, denoted by $F_{gs}$ and $F_{gd}$ respectively, based on whether they are immutable or can be modified by agent actions. An agent's state $\hat{s}$ is described by $\vec{f}\!=\vec{f}_l^i\cup \vec{f}_{gd}\cup\vec{f}_{gs}$.

\begin{definition}
   \textbf{Static global features} $F_{gs}$ are exogenous factors that affect all agents and are observable to all agents but not changed by the agents' actions (e.g. ocean currents).  
\end{definition}

\begin{definition}
   \textbf{Dynamic global features} $F_{gd}$ describe properties of the environment that affect agent operation directly or indirectly, and can be modified by agent's actions (e.g. locations and sizes of the shelves moved by the agents).
\end{definition}
We model the NSE penalty as a function of \emph{dynamic global features} of the joint state, $R_N({s})\!=\!\Omega(\vec{f}_{gd})$, where $\Omega$ is a mapping from $\vec{f}_{gd}$ to the penalty value. This can also be written in the common state-action notation as $R_N(s,a)\!=\!R_N(s), \forall a\!\in\!A$. 

\section{PENALTY DECOMPOSITION}
\label{sec:blame_assgn}
Our algorithm for NSE penalty decomposition, \textit{Reward Estimation using Counterfactual Neighbors (RECON)}, is outlined in Algorithm~\ref{alg:RECON}. RECON first initializes each agent's penalty function $R_N^i$ to zero (Line 1). After the agents have calculated their naive policies independently, the NSE Monitor component in the metareasoner calculates the joint penalty $\vec{r}_N$ for NSEs incurred from joint policy $\vec{\pi}$ (Line 2). If the penalty exceeds a pre-defined NSE tolerance threshold $\eta$, then the Blame Resolver decomposes the joint penalty into local penalties for each agent, based on their relative contribution to NSE, calculated using blame (credit) assignment (Lines 3-5). The penalty decomposition resolves the dependency induced by the joint penalty and enables decentralized policy computation to optimize task completion while minimizing NSEs. 

\begin{algorithm}[t]
\caption{\textsc{RECON}}
\label{alg:RECON}
\textbf{Input}: NSE Tolerance $\eta$, $\vec{\pi}$ for primary assigned task\newline
\begin{algorithmic}[1]
    \STATE Initialize $R^i_N(\hat{s}) = 0, \forall \hat{s}\in \hat{S}_i, \forall i \in \mathcal{A}$
    \STATE Calculate joint penalty $r_N^{\vec{\pi}}$ for $\vec{\pi}$
    \IF{$r_N^{\vec{\pi}}$ $>$ $\eta$}
    \STATE Compute blame $B_i(s)$ using Eqn.~\ref{eq:blame_eq}, $\forall i \in \mathcal{A}, \forall s \in S$
    \STATE $R^{i}_{N}(\hat{s}_i) \gets B_i(s)$, $\forall i \in \mathcal{A}, \forall \hat{s}_i \in s, \forall s \in S$
    \ENDIF
    \RETURN{} $[R^1_N,...,R^m_N]$
\end{algorithmic}
\end{algorithm}

\subsection{Blame Estimation Using Counterfactual Evaluation} A blame value is calculated corresponding to each agent's contribution to the global penalty, using counterfactual information. Since the NSE occurrence is determined only by the dynamic global features $(\vec{f}_{gd})$, counterfactuals must be calculated only over $\vec{f}_{gd}$ to avoid incorrect attribution. For example, consider multiple warehouse robots navigating in a narrow corridor, with some carrying large shelves. The NSE of blocked paths for human access is determined by number of agents carrying large shelves. Generating counterfactuals by considering all state features, including changing agent location, does not provide information on NSE associated with multiple agents carrying large shelves. We therefore estimate blame using \emph{counterfactual neighbors}, which are a subset of counterfactual states.

\begin{definition}The \textbf{counterfactual neighbors} of a joint state $s$, denoted by $s_c$, are the set of all states that differ only in the values of dynamic global features $(\vec{f}_{gd})$ while all other feature values are same as in $s$.
\end{definition}

The set of counterfactual neighbors that are reachable from the start state in the environment are referred to as \emph{valid counterfactual neighbors} and are denoted by ${s}_c^v$. 
\begin{definition}
    \textbf{Agent-specific counterfactual neighbors}, denoted by $s^i_c\!\subset s_c$, are states that differ in those dynamic global features that can be controlled by agent $i$, while other feature values are same as in the current joint state $s$ (e.g. a warehouse robot evaluated with a different sized box).
\end{definition}

Agent-specific counterfactuals estimate an agent's contribution to NSE, similar to credit assignment where the agent contribution is estimated by either removing it or replacing its actions, while keeping others' behaviors fixed~\cite{nguyen2018credit,proper2012modeling}. Let ${s}^{i,v}_{c}\subset s^v_c$ be the set of valid (reachable) counterfactual neighbors for agent $i$.  The blame $B_i$ for agent $i$ in the joint state $s$ is calculated based on the difference between the current NSE penalty and the minimum NSE penalty that could have been achieved by the agent: 

\begin{equation}
    B_i({s}) = \frac{ b_i({s})}{ \sum_{i\in \mathcal{A}} b_i(s)} \cdot {R_N}(s), \text{ with}\label{eq:blame_eq}  
\end{equation}
\begin{equation}
    b_i(s) = \frac{1}{2}\left({R_N^*}+\epsilon+\left(R_N(s) - \min_{s'\in s^{i,v}_{c}} R_N(s')\right)\right),
    \label{eq:blame_calc}
\end{equation}

where $R_N$ is the joint NSE penalty function, $R_N^* = \max_{s\in S}R_N(s)$ denotes the maximum joint NSE penalty possible which is used to rescale blame in the range $[0,R_N^*]$ to avoid impractical values that might be negative or exceed the joint NSE penalty itself. $\epsilon$ is a small fixed value to avoid singularities during normalization and rescaling, and $b_i(s)$ is an intermediate value used to calculate $B_i(s)$. Equation~\ref{eq:blame_eq} assigns blame proportional to the agent's ability to mitigate NSEs. This ensures that agents already making their best efforts are penalized less compared to other agents.

The metareasoner's Blame Resolver compiles a local penalty function $R^{i}_{N}$ for each agent $i$ as follows:
\begin{equation}
    R^{i}_{N}(\hat{s}_i) = B_i(s), \forall \hat{s}_i \in s,\forall s \in S, i\in \mathcal{A}.
    \label{eq:R_blame_eq}
\end{equation}

The agents then solve the MASE-MDP using lexicographic value iteration (LVI)~\cite{wray2015multi}, with the prescribed $R_1^i$ and $R^i_N$ provided by the metareasoner.  

\noindent \textbf{Generalizing }\pmb{$R^i_N$} The penalty function $R^i_N$, calculated using Equation~\ref{eq:R_blame_eq}, is based on the blame values corresponding to $\vec{\pi}$ and does not provide any information about potential NSEs that may occur if the agents followed a \emph{different joint policy}. As a result, NSEs may persist when agents update their policies by solving the MASE-MDP with $R^i_N$. To overcome this limitation, we consider a supervised learning approach to generalize the NSE penalty to unseen situations by using the $R^i_N$, based on the initial $\vec{\pi}$, as training data. The prediction accuracy can further be improved by including the counterfactuals as part of the training~\cite{zerbel2023counterfactual}. Generalization helps scale the algorithm to handle many agents, eliminating the need for an iterative RECON, which incurs a high computational overhead.

\begin{figure*}[t]
    \centering
    \begin{subfigure}[t]{0.30\textwidth}
    \centering
        \includegraphics[width=0.85\linewidth]{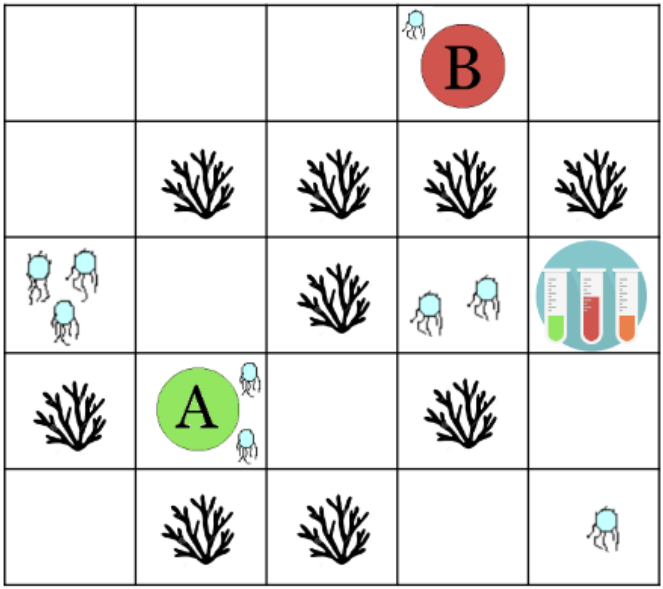}
        \caption{Salp agents \includegraphics[height=2ex]{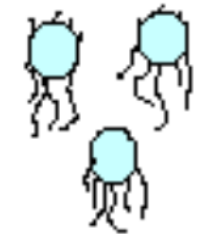}, inspired from \cite{sutherland2010comparative,sutherland2017hydrodynamic}, are tasked with collecting physical samples from sites \includegraphics[height=2ex]{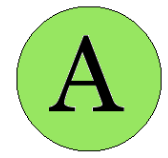},\includegraphics[height=2ex]{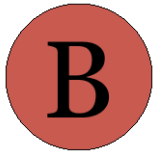}, and depositing them at lab facility \includegraphics[height=2ex]{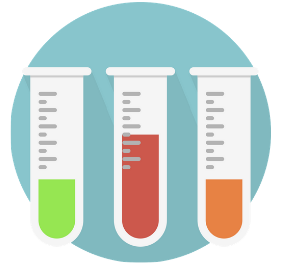} surrounded by corals \includegraphics[height=2ex]{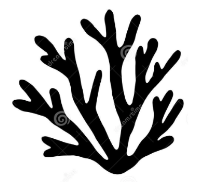} prone to damage.}
        \label{fig:salp_domain}
    \end{subfigure}
    \hfill
    \begin{subfigure}[t]{0.37\textwidth}
    \centering
        \includegraphics[width=0.85\linewidth]{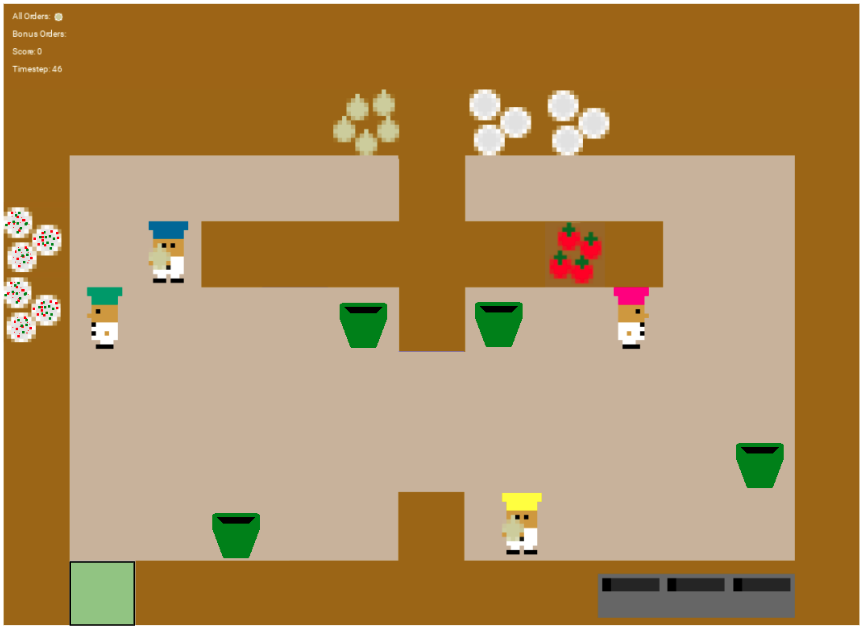}
        \caption{Overcooked environment, inspired from \cite{carroll2019utility}, shows agents cooking and cleaning in kitchen with tomatoes \includegraphics[height=2ex]{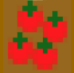}, onions \includegraphics[height=2ex]{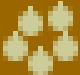}, cooking pots \includegraphics[height=2ex]{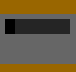}, clean dishes \includegraphics[height=2ex]{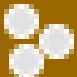}, serving counter \includegraphics[height=2ex]{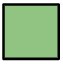}, dirty dishes \includegraphics[height=2ex]{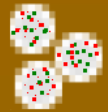}, and waste bins \includegraphics[height=2ex]{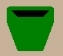}.}
        \label{fig:overcooked_domain}
    \end{subfigure}
    \hfill
    \begin{subfigure}[t]{0.292\textwidth}
    \centering
        \includegraphics[width=0.85\linewidth]{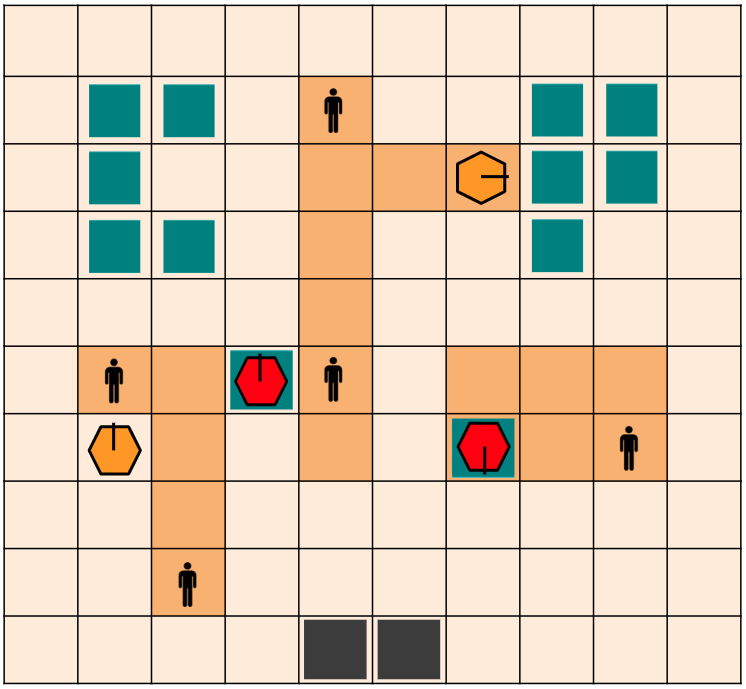}
        \caption{Warehouse environment, inspired from \cite{papoudakis2021benchmarking}, shows agents \includegraphics[height=2ex]{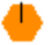} tasked with processing shelves \includegraphics[height=2ex]{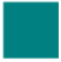}, at counter \includegraphics[height=2ex]{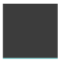}. The narrow corridors \includegraphics[height=2ex]{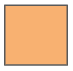} with human workers \includegraphics[height=2ex]{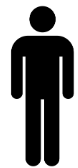}.}
        \label{fig:warehouse_domain}
    \end{subfigure}
    \caption{Instances of environments from (a) salp, (b) overcooked, and (c) warehouse domains used in our experiments.}
    \label{fig:All_domains}
\end{figure*}
\section{EXPERIMENTS}
We evaluate in simulation and using mobile robots. Code will be made public after paper acceptance. 
The MASE-MDP problem is solved using LVI~\cite{wray2015multi}, with zero NSE tolerance $\eta\!=\!0$ and using $\epsilon = 10^{-4}$ for rescaling the blame values.

\begin{figure*}[!t]
    \centering
    \begin{tikzpicture}
        \node[anchor=south] (leg) at (current bounding box.north) {
            \begin{tabular}{c}
                \includegraphics[width=0.95\linewidth]{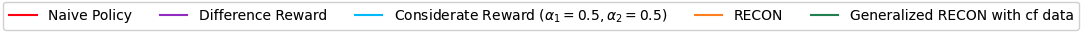}
            \end{tabular}};
    \end{tikzpicture}
    \begin{subfigure}[t]{0.3\textwidth}
        \includegraphics[width=\linewidth]{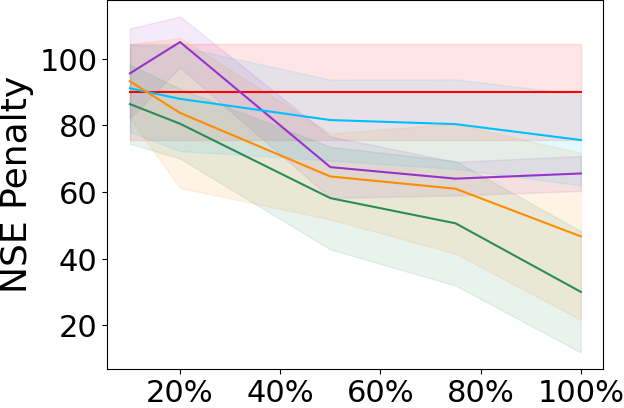}
        \caption{Salp domain}
        \label{fig:salp_corrected}
    \end{subfigure}
    \begin{subfigure}[t]{0.3\textwidth}
        \includegraphics[width=\linewidth]{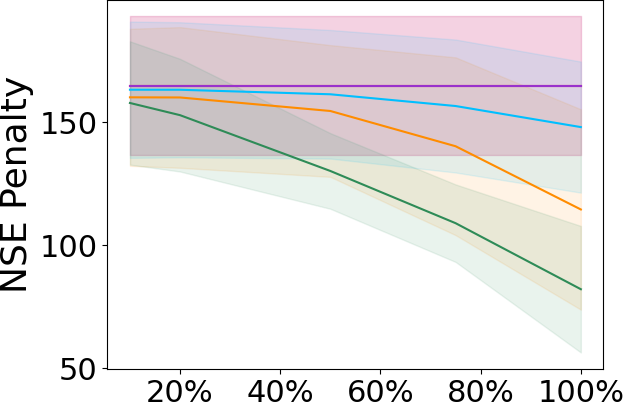}
        \caption{Overcooked domain}
        \label{fig:overcooked_corrected}
    \end{subfigure}
    \begin{subfigure}[t]{0.3\textwidth}
        \includegraphics[width=\linewidth]{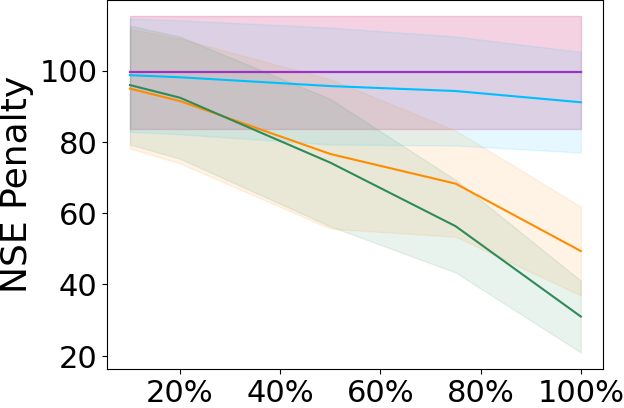}
        \caption{Warehouse domain}
        \label{fig:warehouse_corrected}
    \end{subfigure}
    \caption{Average NSE penalty and standard deviation for varying \% of  agents undergoing policy update in each domain with $25$ agents.}
\label{fig:agents_undergoing_policy_update_all_domains}
\end{figure*}

\begin{figure*}[!t]
    \centering
    \begin{tikzpicture}
        \node[anchor=south] (leg) at (current bounding box.north) {
            \begin{tabular}{c}
                \includegraphics[width=1.02\linewidth]{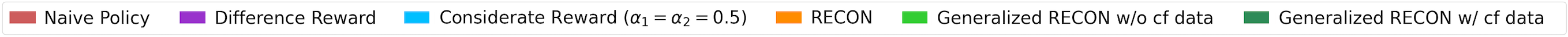} 
            \end{tabular}};        
    \end{tikzpicture}
    \begin{subfigure}[t]{0.3\textwidth}
        \includegraphics[width=\linewidth]{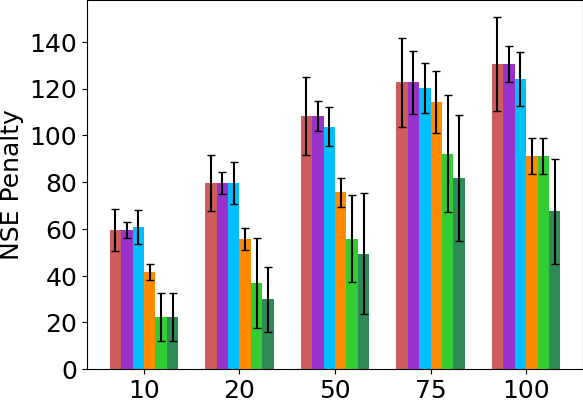}
        \caption{Salp domain}
        \label{fig:salp_generalization}
    \end{subfigure}
    \begin{subfigure}[t]{0.3\textwidth}
        \includegraphics[width=\linewidth]{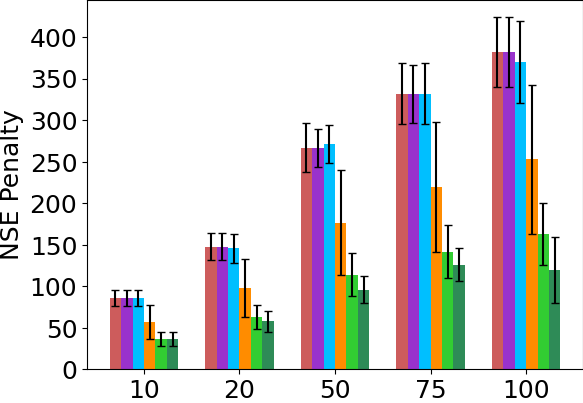}
        \caption{Overcooked domain}
        \label{fig:overcooked_generalization}
    \end{subfigure}
    \begin{subfigure}[t]{0.3\textwidth}
        \includegraphics[width=\linewidth]{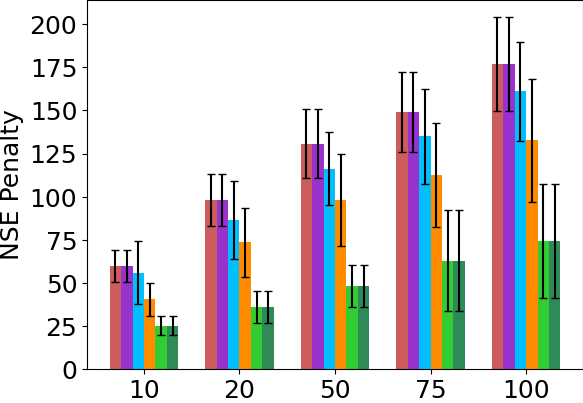}
        \caption{Warehouse domain}
        \label{fig:warehouse_generalization}
    \end{subfigure}
    \caption{Average NSE penalties and standard deviations, averaged over five problem instances with varying number of agents.}
    \label{fig:effect_of_generalization_in_all_domains}
\end{figure*}

\begin{table*}[!t]
  \centering
  \caption{Time taken (in minutes) to solve problems with 100 agents, averaged over five problem instances in each domain.}
   \begin{tabular}{|c|c|c|c|c|c|}
    \hline
    \textbf{Domains}
     & \textbf{Considerate Reward}
     & \textbf{Difference Reward} & \textbf{RECON} & \textbf{Generalized RECON without CF} & \textbf{Generalized RECON with CF} \\
    \hline
    Salp & $1.1\pm 0.1$ & $2.9\pm 0.3$ & $4.0\pm 0.4$ & $5.5\pm 0.8$ & $5.7\pm 0.8$ \\
    \hline
    Overcooked & $121.3\pm 3.8$ & $174.4\pm 3.4$ & $205.1\pm 4.1$ & $254.4 \pm 3.2$ & $288\pm 6.3$ \\
    \hline
    Warehouse & $9.1\pm 0.3$ & $10.0\pm 0.4$ & $12.4\pm 0.2$ & $16.1\pm 0.1$ & $19.0\pm 0.4$ \\
    \hline
  \end{tabular}
  \label{table:scalability_in_all_domains}
\end{table*}
\noindent \textbf{Baselines}  We compare the performance of RECON with: 
\begin{enumerate}
    \item \emph{Naive Policy} that is optimal for the assigned task and does not optimize NSE mitigation, providing an upper bound on NSE penalty;
    \item \emph{Difference Reward}~\cite{proper2012modeling} that has been modified to perform blame assignment only considering dynamic global features, to be consistent with our approach, $B_i(s) = R_N(s) - \max_{s'\in s^{i,v}_{c}}R_N(s')$;
    \item \emph{Considerate Reward} approach~\cite{alizadeh2022considerate} that has been modified to support multiple agents acting simultaneously,
    $ R^i_{r}(\hat{s}_i) = \alpha_1 \frac{R_1^i(\hat{s}_i)}{R_1^*} + \alpha_2 \frac{R_N(s)-B_i(s)}{R_N^*}$ where $R_1^*=\max_{\hat{s}\in \hat {S}_i} R_1(\hat{s})$ is the maximum possible reward for assigned task for agent $i$, $\alpha_1$ and $\alpha_2$ are the selfish and care coefficients respectively. $R_1^*$ and $R_N^*$ are used for normalization so that their relative scales are not an inherent factor, but a controlled one (using $\alpha_1$, $\alpha_2$); 
    \item \emph{Generalized RECON w/o counterfactual data} which generalizes the estimated $R^i_N$ to unseen states, using supervised learning; and 
    \item \emph{Generalized RECON w/ counterfactual (cf) data} which generalizes $R^i_N$ by including the counterfactuals in the training data, using supervised learning.
\end{enumerate}

\noindent \textbf{NSE Penalty Calculation} Our experiments use a logarithmic NSE penalty function to model scenarios where the NSE impact plateaus with a certain number of robots involved~\cite{gemp2022designing,zhu2018joint}, such as negligible difference in penalty between 10 and 11 robots blocking a corridor. The penalty is calculated as, 
\begin{equation}
     R_N(s) = \sum_{d\in F_{gd}}\sum_{k\in d} \beta_{k}\cdot \log\left(\alpha_d N_{k} + 1\right)
\end{equation}
where $d$ is a feature in the set of dynamic global features $F_{gd}$, $N_{k}$ is the number of robots with feature value $d\!=\!k$ in the joint state $s$, and $\beta_k$ is a scaling factor for penalty based on a specific feature value $d\!=\!k$. Essentially, higher NSE penalties correspond to higher values of $\beta_k$, reflecting that addressing key feature values yields a more substantial reduction in NSE penalty. $\alpha_d\, (>0)$ is a sensitivity parameter for NSE associated with feature $d$. A larger $\alpha_d$ denotes that the NSE penalty is more sensitive to an increasing number of robots with a specific feature value. 

\subsection{Evaluation in simulation} Experiments are conducted using the following three domains in simulation and the results are averaged over five instances in each domain.

\noindent \emph{\underline{Sample Collection using Salps}:~}
Salp-inspired~\cite{sutherland2010comparative} underwater robots are tasked with collecting chemical samples of type $A$ or $B$ from different locations in the seabed as illustrated in Fig.~\ref{fig:salp_domain}. 
A robot's state is denoted by $\langle x,y,sample,coral,status\rangle$, where $x,y$ denote robot's location, $sample$ indicates the sample type with $X$ indicating no sample, $coral$ indicates presence of coral at $x,y$, and $status$ indicates if the sample has been deposited at the destination. 
Features used to generalize $R^i_N$ are $\langle sample, coral\rangle$. We test with five $20\!\times\!20$ grids that vary in coral locations. 
\noindent \textbf{NSE:} The salp-like robots may have chemical residues floating around them when transporting samples. A joint NSE occurs when multiple robots carrying chemical samples are in immediate vicinity of corals, potentially damaging it. 

\noindent \emph{\underline{Overcooked}:~} As shown in Fig.~\ref{fig:overcooked_domain}, robots are tasked with preparing a fixed number of tomato and onion soup orders, while keeping the kitchen clean~\cite{zhao2022coordination}. In each problem instance, $20\%$ agents are assigned to cleaning and the rest are assigned cooking tasks. 
An agent's state is represented as $\langle x,y,dir,object,bin,status\rangle$, where $x,y$ denote its location, $dir$ is its orientation, $bin$ indicates the presence of garbage bins at $x,y$, and  task completion status is denoted by $status$. Agents can move forward in all directions, and $interact$ with objects. Interactions include picking up and putting down objects, cooking, and dumping garbage in bins.
Features used for generalizing $R^i_N$ are $\langle object, bin\rangle$. We test with five $15\!\times15$ grids with varying locations of the garbage bins.
\noindent \textbf{NSE:} The garbage bins emit bad odors and attract flies. Any object involved in soup preparation must therefore be kept away from the garbage bins or the waste must be disposed at a farther bin.

\noindent \emph{\underline{Warehouse}:} Robots are assigned specific shelves of different sizes (\emph{big} or \emph{small}) that they need to locate, pick up, transport, process (drop at the counter), and bring back, in order to complete their task~\cite{li2023double,logothetis2021efficient}. A robot's state is represented as $\langle x,y, \emph{shelf size}, \emph{shelf status}, \emph{corridor}, done\rangle$, where $x,y$ is its location, \emph{shelf size} is the size of the shelf transported by the robot and can be one of \emph{big}, \emph{small}, or \emph{X} for no shelf. \emph{shelf status} denotes the processing stage of the assigned shelf which could be one of \emph{\{picked up, processed, delivered\}}, $corridor$ denotes presence of a narrow corridor at a given location $x, y$, and $done$ is a flag indicating task completion.
Features used to generalize $R^i_N$ are $\langle \emph{shelf size},\emph{shelf status}, corridor\rangle$. Test instances, similar to Fig.~\ref{fig:warehouse_domain}, are generated by varying the locations of corridors across the warehouse. 
\textbf{NSE:} When multiple robots carry large shelves simultaneously through the same narrow corridor, it inconveniences human workers trying to access the area. The NSE penalty depends on shelf size and the number of agents carrying shelves in the same corridor. 

\subsection{Results and Discussion}
\noindent \textbf{Number of agents undergoing policy update~}
For each domain in simulation, we consider $25$ agents in the environment and vary the percentage of agents undergoing policy update to minimize NSEs, from $10\%$ to $100\%$ (Fig.~\ref{fig:agents_undergoing_policy_update_all_domains}). We select agents to update policies by ranking them in the decreasing order of their blame values. Note that in some cases, NSE may not be avoided even when we update the policies of $100\%$ of the agents in the system. This is because we prioritize completing the task optimally over minimizing NSEs, and it may be impossible for the agents to avoid NSEs while optimally completing their tasks. This is a problem characteristic and not a limitation of RECON. 
Overall, the results show that RECON and its variants with generalizations can successfully mitigate NSEs, without updating the policies of a large number of agents. 

\noindent \textbf{NSE mitigation and scalability}
Fig.~\ref{fig:effect_of_generalization_in_all_domains} shows that RECON and both version of generalized RECON outperform other methods, consistently reducing NSE penalty across domains and with varying number of agents in the system. The results also show that generalization is useful and can mitigate NSEs considerably even when counterfactual data is not used for training. The above results are with $50\%$ of agents undergoing policy update for each technique, based on the trend in Fig.~\ref{fig:agents_undergoing_policy_update_all_domains}. While updating 100\% agents reduced the NSE penalty in our experiments, it is practically infeasible to implement it in practice for large systems. 

Table~\ref{table:scalability_in_all_domains} shows  the run time (in minutes) of various techniques to solve problems with $100$ agents. The results indicate an approximately linear increase in the run time with increase in the number of agents. 
The considerate reward baseline takes the least time as it solves a single-agent problem with  other agents treated as part of the environment. 

\subsection{Evaluation using mobile robots} 
We conduct experiments in the salp domain with two Turtlebots to validate real-time usability and effectiveness of our approach in mitigating NSEs, using the map in Fig.~\ref{fig:experimental_setup}. The map layout and operation zone of each robot make it inevitable to fully avoid NSEs. Table~\ref{table:hardware_experiments} shows the percentage of NSE states encountered, with standard deviation. We report results only with Generalized RECON w/ counterfactual data as it consistently performs similar to or better than the other RECON versions (Fig.~\ref{fig:effect_of_generalization_in_all_domains}).
\begin{table}[!h]
  \centering
  \caption{Average NSE encounters and standard deviation from our experiments with two mobile robots as shown in Fig.~\ref{fig:experimental_setup}.}
   \begin{tabular}{|c|c|}
    \hline
    \textbf{Approach}
     & \textbf{Average NSE Encounters} \\
    \hline
    Naive & $40.33\%\pm 5.42$ \\
    \hline
    Difference Reward & $38.86\%\pm 6.73$ \\
    \hline
    Considerate Reward & $46.83\%\pm 4.77$ \\
    \hline
    Gen. RECON w/ cf data & $10.33\%\pm 5.33$ \\
    \hline
  \end{tabular}
  \label{table:hardware_experiments}
\end{table}
Considerate Reward with fewer agents is worse than Naive approach, as each agent is less cautious about avoiding NSEs due to reduced joint penalties. Generalized RECON w/ counterfactual data produces the least NSEs, outperforming the baselines. 

\section{SUMMARY AND FUTURE WORK}
This paper formalizes the problem of mitigating NSEs in cooperative multi-agent settings as a decentralized, bi-objective problem. The agents' assigned tasks follow transition and reward independence. The agents produce no NSE when operating in isolation but their joint actions produce NSE and incur a joint penalty.
We present a metareasoning approach to detect NSEs and update agent policies in a decentralized manner, by decomposing the joint NSE penalty into individual penalties. Our algorithm, RECON, uses a counterfactual-based blame attribution to estimate each agent's contribution towards the joint penalty. Our experiments demonstrate the effectiveness of our approach in mitigating NSEs. 
Our framework currently supports Dec-MDPs with transition and reward independence. In the future, we aim to relax this assumption and extend our approach to settings with tightly coupled task assignment. 

\section*{Acknowledgments}
This work was supported in part by ONR grant number N00014-23-1-2171. We thank Akshaya Agarwal and Geoffrey Hollinger for their help with the robot experiments.

\bibliographystyle{IEEEtranS}
\bibliography{references}

\end{document}